\documentclass[runningheads]{llncs}
\usepackage{graphicx}
\usepackage{amssymb}
\usepackage{amsmath}
\begin{document}
\title{A Prior Knowledge Based Tumor and Tumoral Subregion Segmentation Tool for Pediatric Brain Tumors}
% \thanks{Supported by organization x.}
%
\titlerunning{A Prior Knowledge Based Tumor and Tumoral Subregion Segmentation Tool}
% If the paper title is too long for the running head, you can set
% an abbreviated paper title here
%

\author{Silu Zhang \and
Angela Edwards \and
Shubo Wang \and
Zoltan Patay \and
Asim Bag \and
Matthew A. Scoggins}
\authorrunning{S. Zhang et al.}
% First names are abbreviated in the running head.
% If there are more than two authors, 'et al.' is used.
%
\institute{Department of Diagnostic Imaging, St. Jude Children's Research Hospital, Memphis, TN 38105, USA
% \email{\{silu.zhang, angela.edwards, shubo.wang, zoltan.patay, asim.bag, matthew.scoggins\}@stjude.org}
}

% \author{Anonymous\inst{1}}
% \institute{Anonymous Organization
% \email{**@******.***}}

\maketitle              % typeset the header of the contribution
\begin{abstract}
In the past few years, deep learning (DL) models have drawn great attention and shown superior performance on brain tumor and subregion segmentation tasks. However, the success is limited to segmentation of adult's gliomas, where sufficient data have been collected, manually labeled, and published for training DL models. It is still challenging to segment pediatric tumors, because the appearances are different from adult's gliomas. Hence, directly applying a pretained DL model on pediatric data usually generates unacceptable results. Because pediatric data is very limited, both labeled and unlabeled, we present a brain tumor segmentation model that is based on knowledge rather than learning from data. We also provide segmentation of more subregions for super heterogeneous tumor like atypical teratoid/rhabdoid tumor (ATRT). Our proposed approach showed superior performance on both whole tumor and subregion segmentation tasks to DL based models on our pediatric data when training data is not available for transfer learning.
\end{abstract}
\section{Introduction}
Despite the recent advance in deep learning (DL) models on brain tumor segmentation tasks, automatic segmentation of pediatric tumors remains challenging. Pediatric cancer is rare compared to adult cancer and data is complicated by anatomical changes associated with natural development (e.g., changing contrast between grey matter and white matter). The tumor appearance in children is also very different from that in adults. For example, in pediatric high-grade gliomas, both enhancement and edema are much less common than in adult high-grade gliomas~\cite{gutierrez2020radiological}. The majority of existing pretrained deep models were trained on adults' data and perform poorly when applied to pediatric data. The limited data (both labeled and unlabeled) availability of pediatric tumors also hinders opportunities for transfer learning. Further more, for heterogeneous tumors like atypical teratoid/rhabdoid tumor (ATRT), segmenting subregions such as cysts, necrosis, and hemorrhage, in addition to edema and enhancing tumor, is also desired as lesion composition may drive treatment decisions or be related to outcomes. However, most existing models do not provide sub-regioning to this extent. For example, edema and enhancing tumor are the only subregions considered in the BraTS challenges.

Here we describe an automatic brain tumor and subregion segmentation approach that is carefully designed for pediatric tumors including ATRT, diffuse intrinsic pontine glioma (DIPG) and low-grade glioma (LGG). This work has three major characteristics: (1) It is based on extensive use of prior knowledge rather than learning from labeled data, therefore does not require any data for training.
(2) Its performance is better than existing pretrainined DL models and is comparable to a DL model transfer learned on pediatric data, on the whole tumor (WT) segmentation task. 
(3) Our approach also provides cyst, necrosis, hemorrhage, and trapped cerebrospinal fluid (CSF) segmentation in addition to edema and enhancing tumor.

\section{Methods}
\subsection{Overview}
The proposed segmentation approach consists of four major steps: 1) image preprocessing (coregistration, brain extraction, and bias field correction), 2) brain tissue segmentation, 3) WT segmentation, and 4) subregion segmentation. 
The MRI inputs required for segmentation are dependent on the tumor type. For DIPG, and LGG, the algorithm requires T1-weighted (T1), T1 post gadolinium contrast-agent (T1-post), T2-weighted (T2), and T2 fluid-attenuated inversion recovery (FLAIR) images. High cellularity is hallmark of ATRT and the apparent diffusion coefficient (ADC) map is crucial in diagnosis. Therefore, the ADC is included with the 4 above mentioned MRI contrasts as inputs for the segmentation of ATRT. Additionally, we add DWI as input for brain extraction using BET, because a better brain mask can be obtained using DWI than other MRI sequences with the presence of tumor. All input MRIs are coregistered and brain extracted before segmentation. Coregistration is performed using rigid transformation. Bias correction is optional.   
\subsection{Brain tissue segmentation}
\label{sec:brain_tissue_seg}
Traditional and widely used brain tissue segmentation tools such as SPM unified segmentation~\cite{ashburner2005unified} were designed for the healthy population and do not account for tumor and abnormal appearance. Additionally, the performance of DL-based segmentation models heavily depend on the similarity between the training and the test data. Hence, we present a brain tissue segmentation approach which is a modified implementation of the brain tumor segmentation algorithm based on outlier detection~\cite{prastawa2004brain}. Our implementation incorporates significant prior knowledge and is carefully designed and optimized on the pediatric data, including LGG, DIPG and ATRT, with the possibility of adding more tumor types in future. We model brain tissues as four classes: white matter (WM), gray matter (GM), CSF and other, i.e., $\mathbb{Y}=\{\mathrm{WM},\mathrm{GM},\mathrm{CSF},\mathrm{other}\}$. The ``other'' class is a mixture of tumor, blood vessel, abnormal appearance (not labeled as tumor) and non-brain regions (due to imperfect brain extraction). 

\subsubsection{Initialization}
\begin{figure}
\includegraphics[width=\textwidth]{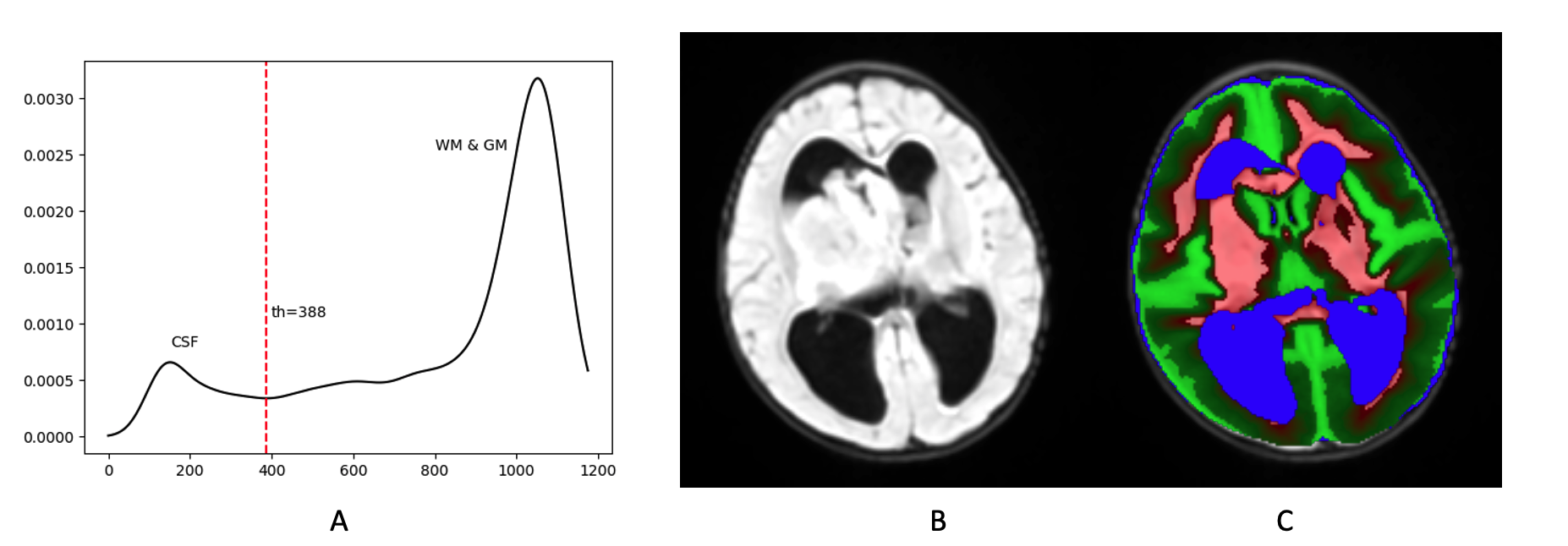}
\caption{An example of prior probability estimation on an ATRT case. A) Estimated probability density function of FLAIR using KDE. B) FLAIR image. C) Estimated prior probabilities of WM (red), GM (green) and CSF (blue).} \label{fig:prior_estimation}
\end{figure}

The WM and GM probabilities are initialized as ICBM atlas priors (affine transformed into the native space) using Advanced Normalization Tools (ANTs~\cite{ants}). Due to abnormal distributions of CSF (e.g., hydrocephalus) common in brain tumor patients, CSF probability is not initialized as the ICBM atlas prior. Our model assumes CSF to be darkest component/peak among all tissue types under consideration in FLAIR. Figure~\ref{fig:prior_estimation}A shows an example histogram from a representative patient with ATRT. The left most peak is CSF and the right most peak is a mixture of WM and GM. The red dotted line is used as the threshold ($th$) for CSF prior estimation, which is determined by the local minima on the right side of the CSF peak. Voxels of intensity value lower than $th$ are initialized as $\Pr(\mathrm{CSF})=0.9$.
Probability of the ``other'' class is initialized as uniform distribution ($
\Pr(\mathrm{other})=0.5$) for non-CSF region.  

\subsubsection{Bayesian inference}
For each input scan $I$, the probability density function of each class $Y$, denoted as $p_I(x|Y)$, is estimated via kernel density estimation using voxels sampled according to prior probability. Because the WM and GM samples can have contaminants (due to inaccurate priors), an outlier detection approach is used to detect and remove the contaminants (key idea proposed in ~\cite{prastawa2004brain}). Outlier detection is performed by using the Minimum Covariance Determinant estimator, setting the support fraction to 0.5. The posterior probabilities are then calculated for each input according to Bayes’ theorem:
$$
{\Pr}_{I}(Y|x)=\frac{p_I(x|Y)\Pr(Y)}{p_I(x)},
$$
where $\Pr(Y)$ is the prior probability, and $p_I(x)=\sum_{Y\in \mathbb{Y}}p_I(x|Y)\Pr(Y).$
The averaged posteriors from all inputs, are used as priors in the next iteration, and the probabilities are updated through a total of 3 iterations in our implementation. The final WM and GM masks are generated by thresholding the smoothed posteriors at 0.5. 

\subsection{Whole tumor segmentation}
\label{sec:WT}
One challenge of tumor segmentation based on abnormal detection is that not everything that has abnormal MRI appearance is tumor. Representative false positives are: vessels (enhanced in T1-post), non-brain regions due to imperfect brain extraction, artifacts (e.g. bias field), and hyperintensities near ventricles due to hydrocephalus. Hence, it is helpful to incorporate prior knowledge with respect to the tumor identification. Our approach aims to segment both homogeneous (e.g., LGG) and heterogeneous (e.g., DIPG, ATRT) tumors. For a heterogeneous tumor, there is no appropriate assumption that can be made on its intensity distribution. However, there is prior knowledge on the MRI appearance of the tumor core for the three tumor types under consideration. ATRTs are hypercellular tumors, therefore are hypointense on ADC. LGG and DIPG both show hyperintense signal on FLAIR. Based on these established characteristics, we detect the WT using a one or two-step method. First, we identify the tumor core for all tumor types.For heterogenous tumors (DIPG and ATRT), we perform an additional step to include surrounding abnormal regions (i.e., edema, necrosis, hemorrhage, and cysts).
\subsection{Subregion segmentation}
\begin{figure}
\includegraphics[width=\textwidth]{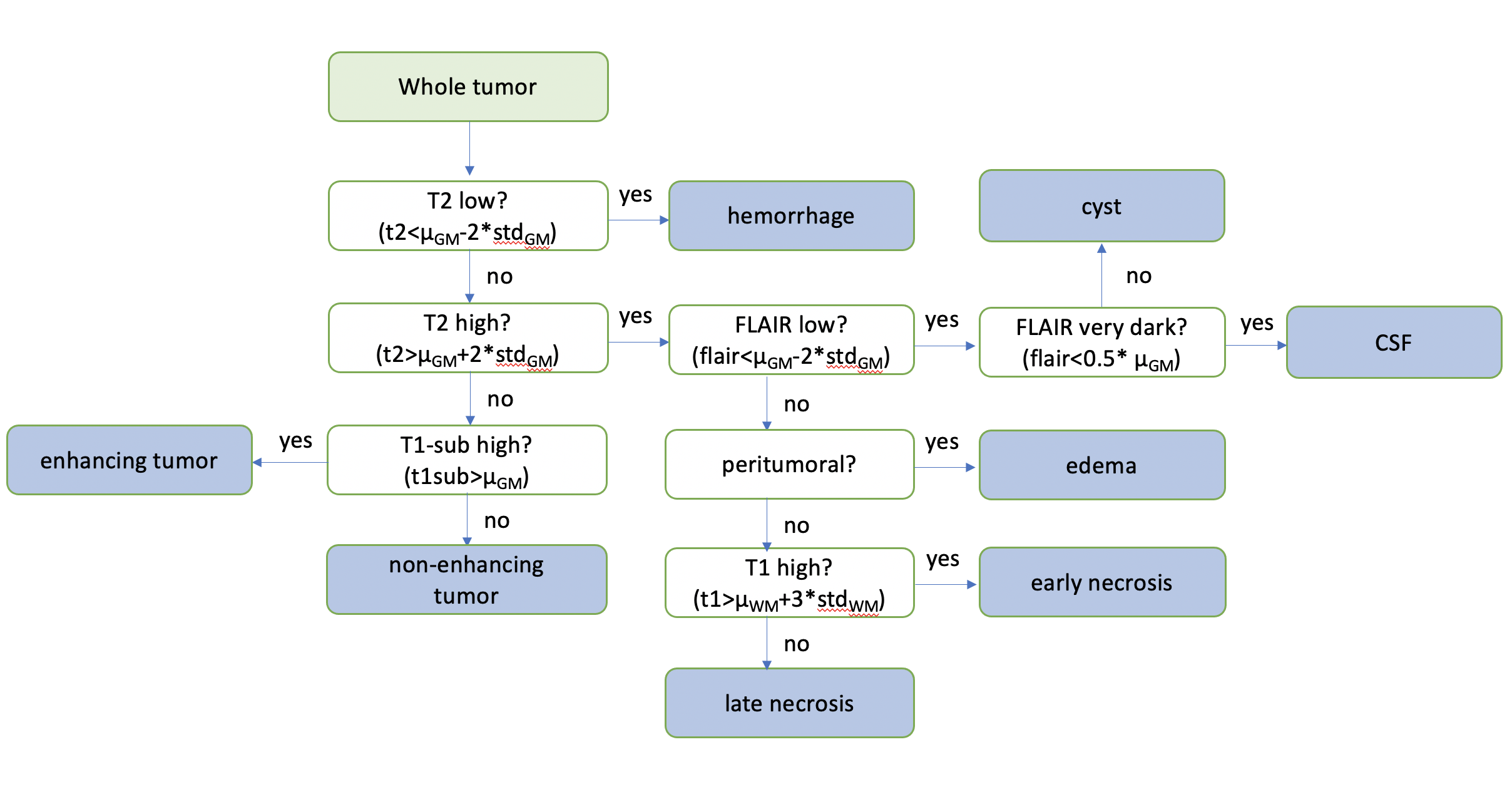}
\caption{Decision rules for subregion segmentations.} \label{fig:DT_subregion}
\end{figure}

Once the WT mask is generated, precise subregions can be obtained by the rules shown in Fig.~\ref{fig:DT_subregion} for heterogeneous tumors (DIPG and ATRT). Subregions include enhancing and non-enhancing tumor (within tumor core), trapped CSF, and the subregions mentioned in Section~\ref{sec:WT}. In Fig.2, the root (green filled) is the WT mask, each leaf node (blue filled) is a subregion, and the splitting criteria are shown in the decision nodes (no fill). Most criteria are based on comparison of intensity to a reference (either WM or GM). Hemorrhage is the dark region in T2. Early and late necrosis are detected as separately because they have different appearance (early necrosis is bright in T1 and late necrosis is bright in T2), even though they are evaluated as a whole to compare with the manual labels. Because edema has similar appearance to late necrosis (both T2 hyperintense), we classify region as edema if it's peritumoral, and late necrosis otherwise. The enhancing tumor is detected from T1-sub, which is the subtraction of T1 from T1-post. CSF and cyst are both detected as dark region in FLAIR, with CSF having a lower threshold than cysts.
\section{Experimental Results}
\subsection{Dataset}
We tested our segmentation tool on three tumor types: ATRT (6 cases), DIPG (4 cases), and LGG (7 cases). WT masks were manually labeled by clinically trained professionals with radiologist oversight. Subregion masks were only manually labeled for ATRT.
\subsection{Methods under comparison}
We compared our proposed segmentation tool with multiple popular pretrained DL models for brain tumor segmentation listed below. All models are publicly available. 
\subsubsection{Segment GBM module from DeepNeuro}
DeepNeuro is a DL-based open-source toolbox for neuroimaging~\cite{beers2021deepneuro}. It provides a module developed for glioblastoma (GBM) edema and enhancing tumor segmentation. This module takes T1, T1-post, and FLAIR as inputs, and produces a WT mask and enhancing tumor mask. The GBM model has a sequential 3D U-Nets structure, which predicts WT mask first and then uses the predicted WT mask as the fourth input to predict enhancing mask~\cite{beers2017sequential}. This model has been trained on both public and private datasets of post-operative GBMs. Because this GBM model has a sequential structure, we were able to modify the implementation to use only the second U-Net to predict enhancing tumor from a given WT mask, for evaluation of enhancing prediction.
\subsubsection{Cascaded Anisotropic Convolutional Neural Networks}
Wang et. al.~\cite{wang2017automatic} developed an  an open-source model that won the 2nd place of MICCAI 2017 BraTS Challenge. The BraTS datasets contain pre-operative multimodal MRI scans of GMB and LGG. The segmentation task is to predict WT, edema, and enhancing tumor given T1, T1-post, T2 and FLAIR. The model consists of three convolutional nerual networks (CNNs) in sequential order. The first CNN (WNet) predicts WT mask, the second CNN (TNet) predicts tumor core (both enhancing and non enhancing tumor) mask within the bounding box defined by the predicted WT mask from the first CNN, then the third CNN (ENet) predicts the enhancing tumor within the bounding box defined by the predicted tumor core mask. Specifically, these CNNs are anisotropic, i.e., each net was trained in axial, sagittal and coronal views respectively using 2D receptive field (a large receptive field in 2D and a relatively small receptive field in the out-plane direction). The authors provided two pretrained models: model15 and model17, trained on BraTS15 and BraTS17, respectively. In addition to these two models, we trained the WNet on our DIPG data (130 cases with manual WT labels) initialized with model17. We label this model as cascaded-CNN-DIPG and add include this method in our comparison for WT segmentation task only. The 4 DIPG cases used for evaluation were not included in the training set for this model. Additionally, we modified the implementation to accept manually labeled WT mask as input and generate subregion masks using TNet and ENet, to evaluate performance of subregion prediction only. 
\subsubsection{NVIDIA Clara}
NVIDIA Clara is an application framework optimized for healthcare and life sciences developers. Clara provides a brain tumor segmentation model based on
autoencoder regularization (clara\_mri\_seg\_brain\_tumors\_br16\_full\_amp)~\cite{myronenko20183d}, which won the 1st place in the BraTS 2018 challenge. The Clara model takes T1, T1-post, T2, and FLAIR as inputs, and outputs WT mask, tumor core mask, and enhancing mask. Unlike other methods under comparison, this model generates WT mask and subregion masks simultaneously and requires GPU to make predictions.
\subsection{Experimental setup}
We evaluated our proposed approach and DL-based models on two separate segmentation tasks: WT segmentation and subregion segmentation. In the WT segmentation task, each method was given the brain extracted images in its desired MR modalities. In the subregion segmentation task, each method was given the manually labeled WT mask as input in addition to the inputs for WT task. Therefore the performance of subregion prediction does not have to be dependent on the WT prediction. However, the NVIDIA Clara model is unable to accept the WT mask as input. As a consequence, its performance does depend on WT prediction. We evaluated subregion segmentation on ATRT, which is typically a heterogeneous tumor. The subregions evaluated are: enhancing tumor, edema, hemorrhage, necrosis (early or late), cyst and CSF. For models that predict WT, tumor core and enhancing tumor, the edema mask was generated using set subtraction of tumor core from WT. A summary of the experimental setup is shown in Table~\ref{table:methods}. The performances of both segmentation tasks were measured by dice score.
\begin{table}
\caption{Experimental setup for methods under comparison.} \label{table:methods}
\includegraphics[width=\textwidth]{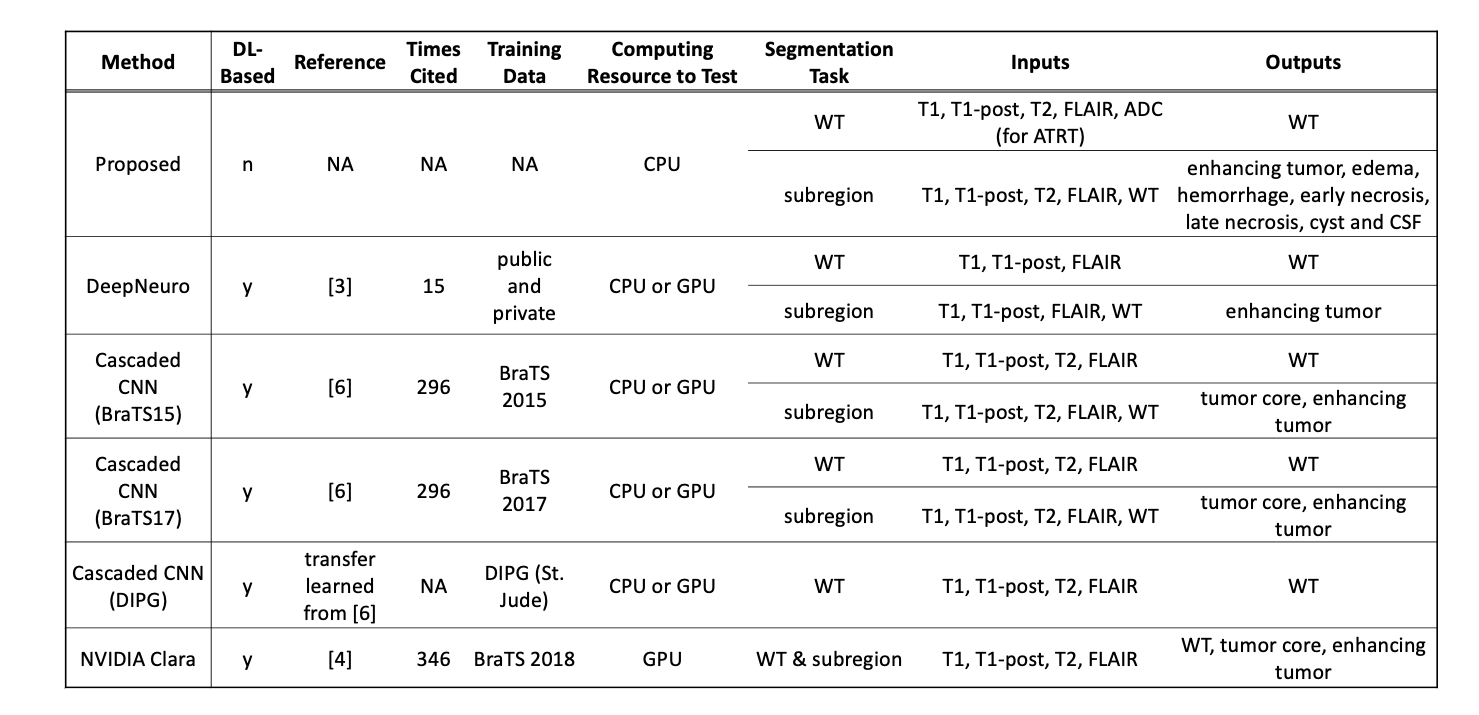}
\end{table}
\subsection{Results}
\subsubsection{WT segmentation}
The WT segmentation results (dice scores) are shown Table~\ref{table:WT}. Our proposed approach outperformed all other DL-based methods on the LGG and ATRT dataset, because the test data is very different from the training data of these DL-based models. For ATRT in particular, our approach had a mean dice score of 0.753, where other methods were below 0.3. All methods consistently performed better on the DIPG tumors compared to the other two data sets. Regarding the DIPG dataset, the best performer was the Cascaded CNN (DIPG) with a mean dice score of 0.924. This was expected because this model has been trained on a sufficient amount of existing DIPG data (130 cases). Our approach performed nearly identical to the 2nd place model, Cascaded CNN (BraTS17), with dice scores of 0.869 and 0.870 respectively, and is substantially  better than the remaining models (0.56, 0.801, and 0.594).

\subsubsection{Subregion segmentation}
\begin{figure}
\includegraphics[width=\textwidth]{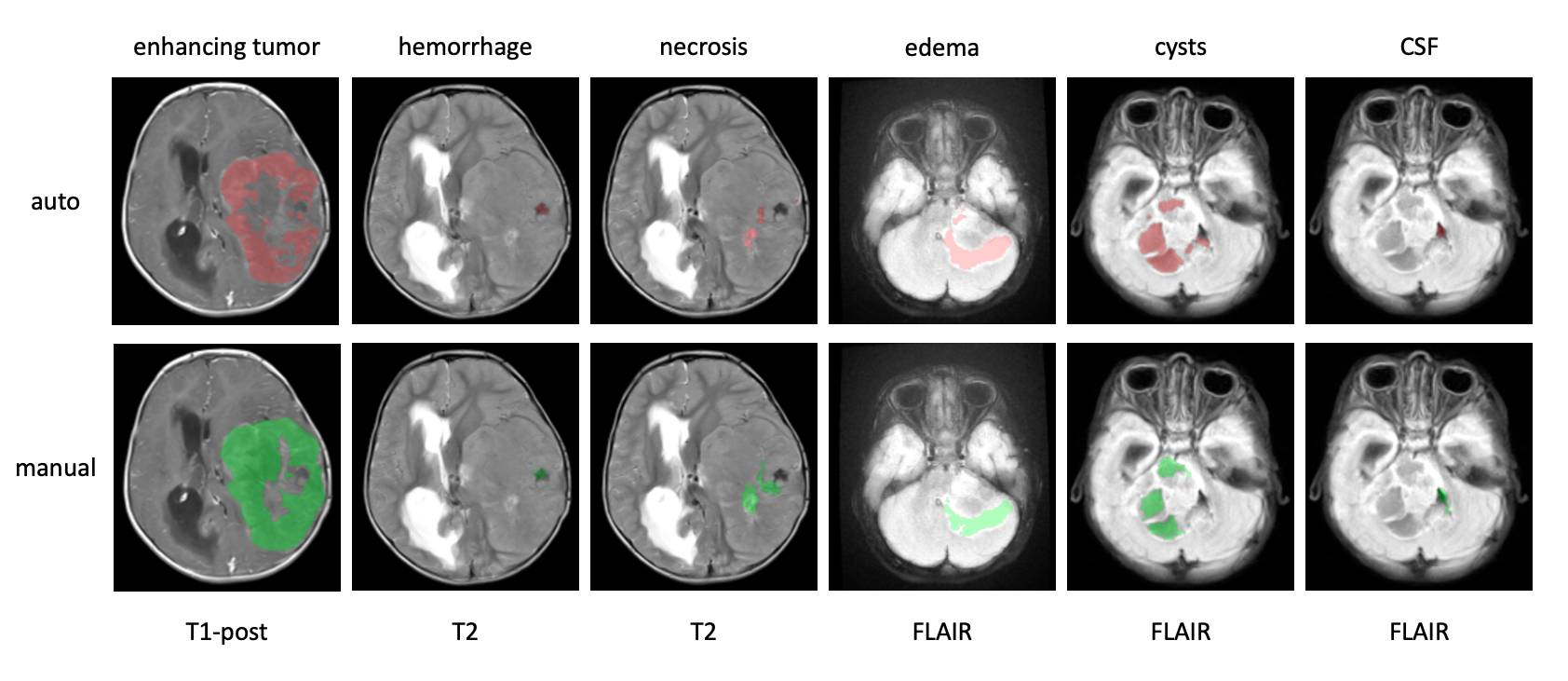}
\caption{Example subregion segmentations of ATRT using the proposed approach. The red ROIs are auto segmentations and the green ROIs are manual segmentations.} \label{fig:subregion_examples}
\end{figure}

Among all subregions under consideration, we were only able to compare our proposed model with others on enhancing tumor and edema segmentations. None of the existing methods used in comparison are capable of making a prediction regarding necrosis, hemorrhage, CSF and cyst, hence we only compared our results with the manual labels. There results of subregion segmentation are shown in Table~\ref{table:subregion}. Our proposed approach had the highest mean dice score on both enhancing and edema segmentation. The results of other subregions (necrosis, hemorrhage, CSF, and cyst) were reasonable and performance varied across specific subregion. Representative examples of subregion segmentations by our approach are shown in Fig. ~\ref{fig:subregion_examples}

\begin{table}
\caption{Performance evaluation (dice score) of all methods under comparison on whole tumor segmentation for different tumor types.} \label{table:WT}
\includegraphics[width=\textwidth]{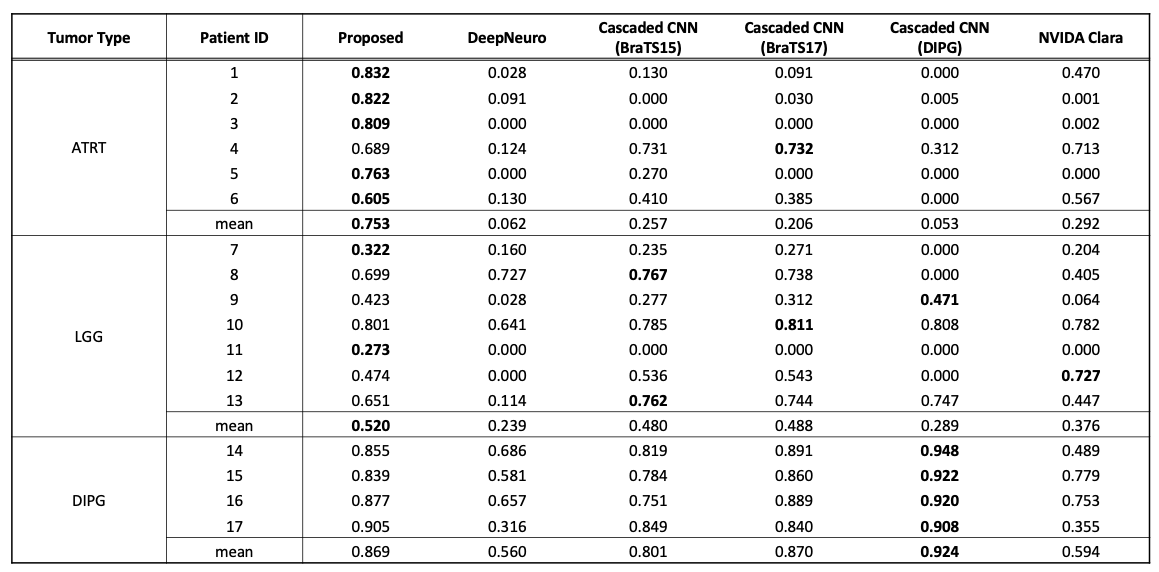}
\end{table}

\begin{table}
\caption{Performance evaluation (dice score) of all methods under comparison on subregion segmentation of ATRT.} \label{table:subregion}
\includegraphics[width=\textwidth]{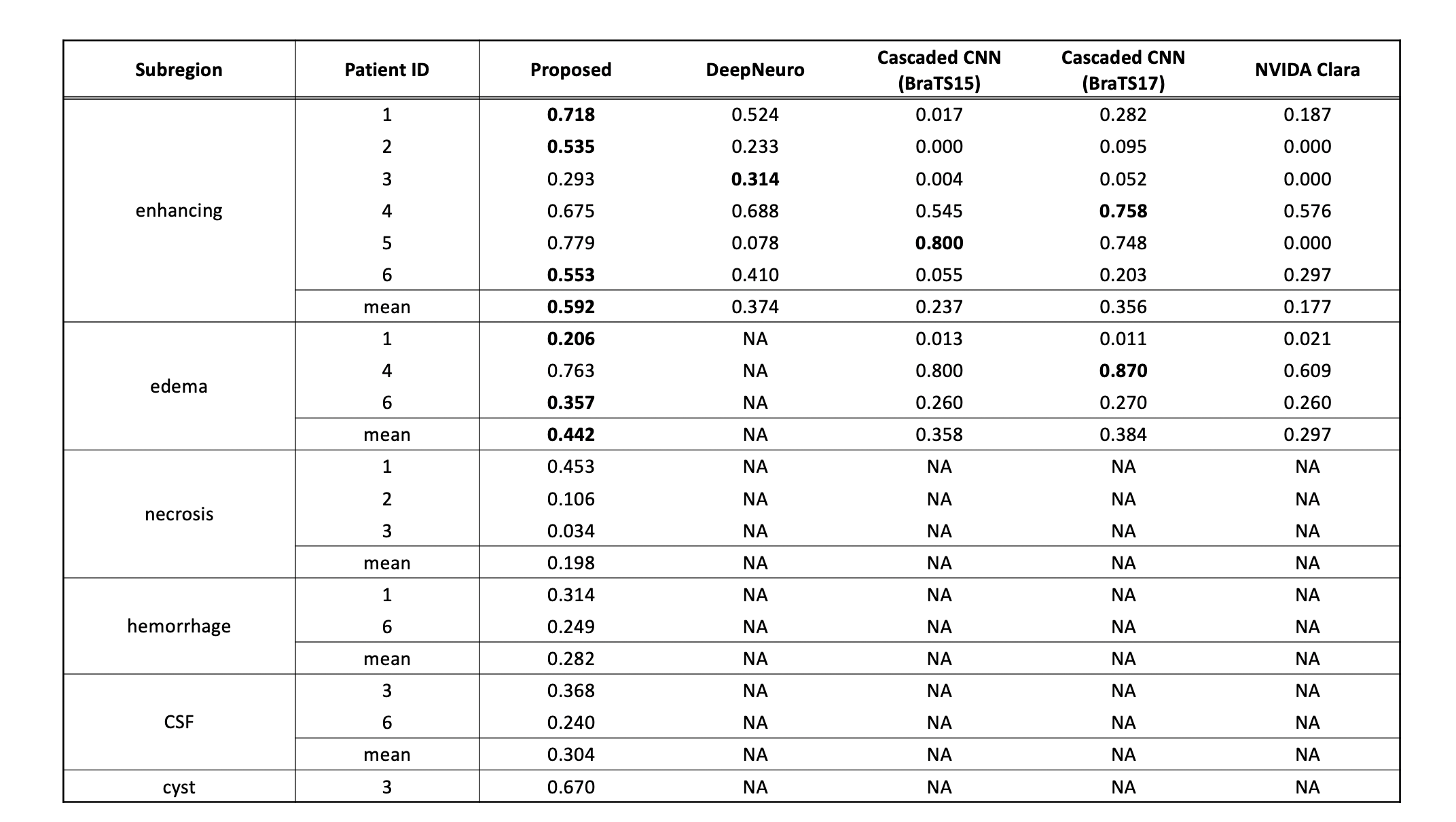}
\end{table}

\section{Discussions}
Automatic segmentation of pediatric tumors is challenging. Our technique offers an approach to automatically segmenting rare pediatric brain tumors that performs consistently and superior to publicly available brain tumors models. Additionally, our knowledge-based brain tumor segmentation approach has shown many advantages over DL-based methods:
\begin{itemize}
	\item More efficient. Knowledge of tumor can be transferred into the model using our approach than using labeled data to train. This is intuitive, for example we can easily detected enhancing tumor from T1-sub by thresholding. However, it may take more than 100 training examples for the model to learn this simple knowledge. A big drawback of DL models is that it is very hard to incorporate prior knowledge, and hence requires large training data to compensate for that. Pediatric cancer is rare compared to adult disease and lack publicly available data sets. 
	\item More generalizable. Our approach can be designed for different tumor types accounting for their similarities and variabilities. For example, we added ADC as an additional inputs for ATRT segmentation, while for LGG and DIPG we only used T1, T1-post, T2, and FLAIR. We added surrounding tumor-related regions for ATRT and DIPG, but not for LGG. However, for DL models to handle different tumor types, separate models for each tumor type or complex architectures are required.
	\item More interpretable. Our approach generates consistent and informative intermediate results for quality checking. Even with the push for explainable AI, DL models can still be difficult to understand, particularly in the case of misclassifications. Our technique is straightforward. For example, our approach estimates WM, GM, and CSF probabilities before generating a WT mask. Determining the reason behind a misclassification is uncomplicated. Additionally, the decision rules for subregion detection are very interpretable and can be easily modified for different tumor types if necessary.
\end{itemize}
In general, DL applications will outperform traditional analytical techniques like the one we propose here. However, performance is determined by the training data. As stated previously, pediatric cancer is rare and access to data (labeled and unlabeled) is limited. The presence of labeled data is increasing, however, through consortia and federated learning. Until public databases are curated and available to facilitate DL models, our approach provides a starting point for assisted annotation. A priori knowledge is a powerful tool and is especially useful for autosegmenting rare tumors like ATRT.

\section{Conclusions}
In this paper, we present a knowledge-based algorithm to segment whole brain tumor and associated subregions. Our approach was customized for different pediatric brain tumors where available labeled data is very limited and far from sufficient to train a learning-based model. Our proposed approach showed superior performance to existing pretrained DL-based models on our ATRT and LGG dataset. On the DIPG dataset, it showed comparable performance to a DL model that have been transfer learned on our DIPG data.      

% \section*{Acknowledgements} 
%
% ---- Bibliography ----
%
% BibTeX users should specify bibliography style 'splncs04'.
% References will then be sorted and formatted in the correct style.
%
\bibliographystyle{splncs04}
\bibliography{references}

\begin{thebibliography}{1}
\providecommand{\url}[1]{\texttt{#1}}
\providecommand{\urlprefix}{URL }
\providecommand{\doi}[1]{https://doi.org/#1}

\bibitem{ashburner2005unified}
Ashburner, J., Friston, K.J.: Unified segmentation. Neuroimage  \textbf{26}(3),
   839--851 (2005)

\bibitem{ants}
Avants, B.B., Tustison, N.J., Johnson, H.J.: Advanced Normalization Tools,
  \url{http://stnava.github.io/ANTs/}

\bibitem{beers2021deepneuro}
Beers, A., Brown, J., Chang, K., Hoebel, K., Patel, J., Ly, K.I., Tolaney,
  S.M., Brastianos, P., Rosen, B., Gerstner, E.R., et~al.: Deepneuro: an
  open-source deep learning toolbox for neuroimaging. Neuroinformatics
  \textbf{19}(1),  127--140 (2021)

\bibitem{beers2017sequential}
Beers, A., Chang, K., Brown, J., Sartor, E., Mammen, C., Gerstner, E., Rosen,
  B., Kalpathy-Cramer, J.: Sequential 3d u-nets for biologically-informed brain
  tumor segmentation. arXiv preprint arXiv:1709.02967  (2017)

\bibitem{gutierrez2020radiological}
Gutierrez, D.R., Jones, C., Varlet, P., Mackay, A., Warren, D., Warmuth-Metz,
  M., Aliaga, E.S., Calmon, R., Hargrave, D.R., Ca{\~n}ete, A., et~al.:
  Radiological evaluation of newly diagnosed non-brainstem pediatric high-grade
  glioma in the herby phase ii trial. Clinical Cancer Research  \textbf{26}(8),
   1856--1865 (2020)

\bibitem{myronenko20183d}
Myronenko, A.: 3d mri brain tumor segmentation using autoencoder
  regularization. In: International MICCAI Brainlesion Workshop. pp. 311--320.
  Springer (2018)

\bibitem{prastawa2004brain}
Prastawa, M., Bullitt, E., Ho, S., Gerig, G.: A brain tumor segmentation
  framework based on outlier detection. Medical image analysis  \textbf{8}(3),
  275--283 (2004)

\bibitem{wang2017automatic}
Wang, G., Li, W., Ourselin, S., Vercauteren, T.: Automatic brain tumor
  segmentation using cascaded anisotropic convolutional neural networks. In:
  International MICCAI brainlesion workshop. pp. 178--190. Springer (2017)

\end{thebibliography}

% \begin{thebibliography}{8}
% \bibitem{ref_article1}
% Author, F.: Article title. Journal \textbf{2}(5), 99--110 (2016)

% \bibitem{ref_lncs1}
% Author, F., Author, S.: Title of a proceedings paper. In: Editor,
% F., Editor, S. (eds.) CONFERENCE 2016, LNCS, vol. 9999, pp. 1--13.
% Springer, Heidelberg (2016). \doi{10.10007/1234567890}

% \bibitem{ref_book1}
% Author, F., Author, S., Author, T.: Book title. 2nd edn. Publisher,
% Location (1999)

% \bibitem{ref_proc1}
% Author, A.-B.: Contribution title. In: 9th International Proceedings
% on Proceedings, pp. 1--2. Publisher, Location (2010)

% \bibitem{ref_url1}
% LNCS Homepage, \url{http://www.springer.com/lncs}. Last accessed 4
% Oct 2017
% \end{thebibliography}
\end{document}